\documentstyle{article}

\begin{document}
\hfill SINP/TNP/93-23\break\smallskip
\hfill hep-th/9312128\break
\bigskip
\begin{center}
\large{\bf STRUCTURE OF MULTI-ANYON WAVEFUNCTIONS}
\end{center}
\bigskip\bigskip\bigskip
\begin{center}
P. Mitra\footnote{e-mail address mitra@saha.ernet.in}\\
Saha Institute of Nuclear Physics\\
Block AF, Bidhannagar\\
Calcutta- 700 064, INDIA
\end{center}
\bigskip\bigskip\bigskip
\begin{center}
{\bf Abstract}
\end{center}

Symmetries  of  multi-anyon  wavefunctions  are analysed with the
help of a  second  quantized  formulation.  Analogues  of  Slater
determinants   are  constructed.  It  is  shown  that  the  Pauli
principle is not enforced.

\vfill \eject

The  description  of  states  containing  two  or  more identical
particles  is  a  standard  part  of  quantum   mechanics   books
\cite{book}.  It  has  been  well  known for a long time now that
wavefunctions of such states behave in one of two possible  ways:
they  are  either  symmetric  or  antisymmetric under exchange of
coordinates of the particles. The two cases  correspond  to  Bose
and  Fermi  statistics  respectively.  But  if  one  considers  a
spacetime which is (2+1) dimensional, the situation becomes  more
complicated \cite{0}. There is no sharp division into two classes
here; one can have a continuous range of statistics, often called
{\it  anyon}  statistics.  Anyons,  which  exist  only  in  (2+1)
dimensions, interpolate between Bose and Fermi statistics.

Although they exist in a somewhat exotic  spacetime,  anyons  are
likely  to  be  physically  important, with significant condensed
matter applications. That is why this  subject  has  been  keenly
followed  in  the  recent  literature  \cite{rev}. Quite a lot is
known already about the quantum mechanics  of  anyons  \cite{qm}.
Some  progress  has  also  been made in the field theory of these
objects \cite{ft}.  However,  the  behaviour  under  exchange  of
particle  coordinates  of  multianyon  wavefunctions has not been
properly studied in the light of the field theoretic formulation.
This is what is proposed to be done in this note.

When a particle is taken round an identical particle, it picks up
a factor which is $\pm 1$ for bosons and fermions, and some phase
factor   for   anyons.  It  follows  that  instead  of  having  a
wavefunction  that  is  symmetric  or  antisymmetric  under   the
exchange  of  coordinates,  two  or  more  anyons can only have a
wavefunction which changes by a phase under such an exchange. The
modulus of  the  wavefunction  must  remain  unaltered.  But  the
modulus  is  not  a  linear function of the wavefunction, so this
statement  is  not  particularly  useful,  and  a  more  detailed
description  is  necessary.  The  phase factor associated with an
exchange of identical anyons may be taken to be  $q=e^{i\theta}$.
The  parameter  $q$ or $\theta$ will of course depend on the type
of anyon considered. For  instance,  in  the  limiting  cases  of
bosons  and  fermions,  $\theta$  can  be taken to be 0 or 1. But
there is an  extra  complication  for  anyons.  In  general,  the
inverse  of  $q=e^{i\theta}$  is  unequal  to  $q$,  so  that the
exchange   has   to   be   assigned   a   direction   or   sense.
Correspondingly,   the   transformation  group  that  enters  the
discussion is not the  permutation  group  but  the  {\it  braid}
group. We shall have more to say about these groups later. Let us
note  now  that  in  the  literature, the dependence of the phase
factor on the direction of the exchange is taken into account  by
specifying  whether  the  rotation is clockwise or anticlockwise.
This corresponds to the  use  of  multivalued  wavefunctions.  We
prefer an alternative description where a {\it cut} is chosen and
wavefunctions  become  single  -valued  but  discontinuous. Then,
instead of writing simply  $q$,  it  is  necessary  to  introduce
$q(\vec  x,\vec  y)$, which is equal to $q$ or $q^{-1}$ depending
on the relative positions $\vec x, \vec y$.

This  is  best  understood  by  referring to the second quantized
picture and writing
\begin{equation}
a^\dagger  (\vec  x)  a^\dagger  (\vec  y)  =  q(\vec  x,\vec  y)
a^\dagger (\vec y)a^\dagger (\vec x), \label{cr}
\end{equation}
where creation operators have been taken to exist. It is believed
that  there has to be some amount of nonlocality in theories with
anyons. But this nonlocality does not manifest itself  in  simple
commutation  relations  like  this,  which  are familiar in model
field theories containing anyonic  excitations.  Self-consistency
demands that
\begin{equation}
q(\vec x,\vec y) = q(\vec y,\vec x)^{-1}. \label{cons}
\end{equation}
In  the  usual  situation  of  Bose  or  Fermi  statistics, where
$q^2=1$, $q$ is independent of location and therefore continuous.
But  in  the  case  of  anyons  with  $q^2\neq  1$,  $q$   is   a
discontinuous function of its arguments. One possibility that can
be visualized for the cut is such that
\begin{equation}
q(\vec  x, \vec y) = e^{i\theta\epsilon (\alpha x_1 + \beta x_2 -
\alpha y_1 - \beta y_2)},
\end{equation}
where  $\epsilon$  represents  the  sign  function. This is self-
consistent but discontinuous. We shall not need the explicit form
of  $q(\vec  x, \vec y)$.

In  general, by  translation  invariance, $q(\vec x, \vec y)$ can
depend only on the difference $\vec x -  \vec  y$.  Further,  eq.
(\ref{cons}) leads to
\begin{equation}
q(\vec  x, \vec x) = \pm 1. \label{q0}
\end{equation}
These  two  possibilities can be referred to as the cases of {\it
bosonic} anyons and {\it fermionic} anyons respectively, if there
is  no  confusion.  In the literature, {\it ferm-ionic} anyons are
predominant. It should be emphasized that  this  distinction  and
this  nomenclature  refer  exclusively  to  the behaviour at zero
separation and are not of fundamental importance.

We  come  now to the behaviour of 2-anyon wavefunctions under the
exchange of particle coordinates. A wavefunction $\psi  (\vec  x,
\vec y)$ corresponds to the 2-anyon creation operator
\begin{equation}
\int  d\vec  x\int  d\vec  y  \psi (\vec  x,  \vec  y)  a^\dagger
(\vec x) a^\dagger (\vec y).
\end{equation}
Now $\psi$ can be decomposed uniquely into  what  may  be  called
its {\it q-symmetric} and {\it q-antisymmetric} parts:
\begin{eqnarray}
\psi(\vec  x,  \vec  y)  &=&  \psi_{q~sym}  (\vec  x,  \vec  y) +
\psi_{q~ant}  (\vec  x, \vec  y)\nonumber\\  &\equiv  &   {1\over
2}[\psi(\vec x, \vec y) + q(\vec y, \vec x) \psi(\vec y, \vec x)]
+  {1\over 2} [\psi(\vec x, \vec y) - q(\vec y, \vec x) \psi(\vec
y, \vec x)]. \label{sym}
\end{eqnarray}
The above creation operator can therefore be written as
\begin{equation}
\int  d\vec  x\int  d\vec  y  (\psi_{q~sym}  (\vec  x,  \vec y) +
\psi_{q~ant}  (\vec x, \vec y))a^\dagger (\vec x) a^\dagger (\vec
y).
\end{equation}
Now by interchanging the integration variables $\vec x$ and $\vec
y$, and making use of eq. (\ref{cr}), one sees that
\begin{eqnarray}
& &\int   d\vec  x\int  d\vec  y ~\psi_{q~ant}  (\vec  x, \vec y)
a^\dagger (\vec x) a^\dagger (\vec y)\nonumber \\ &=& \int  d\vec
x\int d\vec y {1\over 2} \psi(\vec x, \vec y) [a^\dagger (\vec x)
a^\dagger  (\vec  y)  -  q(\vec  x,  \vec  y)  a^\dagger (\vec y)
a^\dagger (\vec x)] \nonumber\\ &=& 0.
\end{eqnarray}
This shows that the {\it q-antisymmetric} part of $\psi$ does not
contribute  to  this  operator, so that $\psi$ can be taken to be
{\it  q-symmetric}  without  loss   of   generality.   Thus   the
generalization  of  the symmetry or antisymmetry of two- particle
wavefunctions  to  the  anyonic  case  is  the  {\it  q-symmetry}
introduced in eq. (\ref{sym}).

The  reader  should be warned to be careful with the order of the
arguments in  these  definitions.   With  the  convention  chosen
above,  the  product  $a^\dagger  (\vec x) a^\dagger (\vec y)$ is
{\it not q-symmetric}. Indeed, eq. (\ref{sym}) shows that
\begin{equation}
\psi_{q~sym}  (\vec  x, \vec y) = q (\vec y, \vec x) \psi_{q~sym}
(\vec y, \vec x),
\end{equation}
whereas eq. (\ref{cr}) has $q (\vec x, \vec y)$ in it. Since  the
interchange  of the two arguments of $q$ amounts to inverting it,
one  concludes that $a^\dagger (\vec x) a^\dagger (\vec y)$ is in
fact {\it q$^{-1}$-symmetric}. The  lesson  is  that  {\it  if  a
q$^{-1}$-symmetric  object  is  contracted  with  anything,  that
object  can  be  taken  to  be  q-symmetric   without   loss   of
generality.} For $q=1$, this is of course a well-known result.

The  generalization  to  states of more than two particles is not
difficult, but there  are  some  extra  complications.  To  study
3-anyon  states,  note  that  the  product  $a^\dagger  (\vec  x)
a^\dagger    (\vec    y)    a^\dagger     (\vec   z)$   is   {\it
q$^{-1}$-symmetric} in $\vec x$ and $\vec y$, and also  in  $\vec
y$  and  $\vec  z$,  but  there  is  no  simple symmetry property
corresponding to the  interchange  of  $\vec  x$  and  $\vec  z$.
However,  this  interchange  can be written in terms of the other
two interchanges, so that a formula can be constructed to express
the  symmetry  property  corresponding  to  this interchange. The
real question is about the wavefunction. By virtue of  the  above
proposition regarding {\it q-symmetric} objects, it is clear that
the   wavefunction  $\psi(\vec  x,  \vec  y,  \vec  z)$  is  {\it
q-symmetric} in its first two arguments and also in its last  two
arguments:
\begin{eqnarray}
\psi(\vec  x, \vec y, \vec z) &=& q (\vec y, \vec x) \psi
(\vec y, \vec x, \vec z)\nonumber\\&=& q (\vec z, \vec y) \psi
(\vec x, \vec z, \vec y).
\end{eqnarray}
{}From this it follows that
\begin{equation}
\psi(\vec x, \vec y, \vec z) = q (\vec y, \vec x) q (\vec z, \vec
x) q (\vec z, \vec y)\psi (\vec z, \vec y, \vec x).
\end{equation}
Although  the  exchange of the first and the third arguments does
amount  to  multiplying  the wavefunction by a phase factor, {\it
this phase factor depends on the second argument too.} This  kind
of  relation  can  also be found for wavefunctions of states with
even more anyons.

These  symmetries  can  be   used   to   construct   many  -anyon
wavefunctions  starting  from those for single anyons. If $\psi_1
(\vec x)$ and $\psi_2 (\vec x)$ are  two  such  wavefunctions,  a
"noninteracting"  2-anyon  wavefunction may in principle be found
by   multiplying   these   functions   and   appropriately   {\it
q-symmetrizing}   the   product.   It  is  assumed  here  that  a
Hamiltonian constructed by adding two single -anyon  Hamiltonians
without   any  interaction  gives  a  reasonable  description  of
reality. From eq. (\ref{sym}), the wavefunction is found to be
\begin{equation}
{1  \over  2}  [\psi_1(\vec x) \psi_2(\vec y) + q(\vec y, \vec x)
\psi_1 (\vec y)\psi_2(\vec x)].
\end{equation}
This expression has two terms corresponding to the  two  ways  of
permuting  the particles. However, unlike the usual cases of Bose
or Fermi statistics, the superposition involves  the  nonconstant
function   $q$.    For  $q=\pm  1$,  the  usual  symmetrized  and
antisymmetrized expressions are of  course  obtained.  Since  for
$q=-1$ a determinant appears, one can, in the general case, speak
of a {\it (-q) determinant} (apart from a factor of a half).

The  3-anyon  wavefunction is more complicated. In terms of three
single -anyon wavefunctions, one can construct the  appropriately
{\it q-symmetrized} product
\begin{eqnarray}
& &{1\over 3!}[\psi_1(\vec x) \psi_2(\vec y) \psi_3(\vec z) + q(\vec
y,  \vec x) \psi_1(\vec y) \psi_2(\vec x) \psi_3(\vec z) + q(\vec
z,   \vec   y)  \psi_1(\vec  x)  \psi_2(\vec  z)  \psi_3(\vec  y)
\nonumber\\ &+& q(\vec z, \vec y) q(\vec z, \vec  x)  \psi_1(\vec
z)  \psi_2(\vec  x)  \psi_3(\vec y) + q(\vec y, \vec x) q(\vec z,
\vec x) \psi_1(\vec y) \psi_2(\vec z) \psi_3(\vec x)  \nonumber\\
&+&  q(\vec  y,  \vec  x)  q(\vec  z,  \vec  x) q(\vec z, \vec y)
\psi_1(\vec z) \psi_2(\vec y) \psi_3(\vec x)]
\end{eqnarray}
There  are  six  terms  here  corresponding  to  the  six ways of
permuting the particles, and  once  again  there  are  nontrivial
location -dependent coefficients for these terms. When $q=1$, the
product is simply the symmetrized product, and  for  $q=-1$,  the
usual determinant appears.  In  general  one  can  refer  to  the
expression in the square brackets as a  {\it  (-q)  determinant}.
{}From   these   two   examples,  the  construction  of  {\it  (-q)
determinants} for an arbitrary number of identical anyons  should
be clear.

A  comment  about the Pauli principle is in order. In the case of
fermions, the fact that $q=-1$ leads to  the  vanishing  of  a  2
-particle wavefunction for coincident arguments,
\begin{eqnarray}
\psi(\vec  x, \vec x) &=& - \psi(\vec x, \vec x) \nonumber \\ &=&
0,
\end{eqnarray}
and  this argument is thought to generalize to the case of anyons
where $q$ is not in general equal to unity.  However,  as  argued
above  in eq. (\ref{q0}), the {\it relevant} value of $q$ is $\pm
1$ depending on whether the anyons are of the  {\it  bosonic}  or
the  {\it fermionic} type,  so this generalization holds only for
the latter type. There is no need for  the  wavefunction  of  two
{\it  bosonic}  anyons  to  vanish when the coordinates coincide.
More important,  the  {\it  q-symmetrized}  products  constructed
above  are  in  general nonvanishing even when two or more of the
single -anyon wavefunctions coincide, and this is true  for  both
{\it  fermionic}  and  {\it bosonic} anyons. This proves that the
Pauli exclusion principle does {\it not} hold in the general case
of anyons and is restricted to the case of fermions for which  it
was after all intended.

Lastly,  some  comments have to be made about the supersession of
the permutation group by the braid group in the case  of  anyons.
For  usual  statistics,  a  multiparticle wavefunction changes by
$\pm 1$ under an interchange of particles, the sign factor  being
related  to  the  statistics  and  the  permutation involved. For
bosons, all permutations give rise to the factor $+1$, while  for
fermions,  the  sign  factor is $(-1)^P$, where $P$ is the parity
(even or odd) of the permutation. In both cases, the  permutation
group  is  represented  one  -dimensionally  on the multiparticle
wavefunction. In the case of anyons,  the  factors  which  appear
when  two  particles  are interchanged are phase factors like $q$
and their products. In general, these phase factors do {\it  not}
form  a  group.  To  have  a  group structure, squares and higher
powers of such phase factors would have to be accommodated.  Such
a  situation  could arise if multivalued wavefunctions were used.
But  we  have  preferred  to  introduce   cuts   and   stick   to
(discontinuous) single- valued wavefunctions. A representation of
the braid group is {\it not} obtained.

It is a pleasure to thank Avinash Khare  for  asking  a  question
and thereby initiating this analysis.

\end{document}